\documentclass{iaus}
\usepackage{graphicx}

\title[Modeling the dust SED of NGC 4214]{Modeling the dust Spectral Energy Distribution of NGC 4214}

\author[Hermelo et al.]   %% give here short author list %%
{Israel Hermelo$^1$, Ute Lisenfeld$^1$, M\'onica Rela\~no$^1$, Richard Tuffs$^2$, Joerg Fischera$^3$, Brent Groves$^4$ and Cristina Popescu$^5$}

\affiliation{$^1$Departamento de F\'isica Te\'orica y del Cosmos, Universidad de Granada, Spain\\
                 email: {\tt israelhermelo@ugr.es, ute@ugr.es, mrelano@ugr.es} \\[\affilskip]
             $^2$Max Planck Institut f\"ur Kernphysik, Heidelberg, Germany\\
                 email: {\tt Richard.Tuffs@mpi-hd.mpg.de} \\[\affilskip]
             $^3$Canadian Institute for Theoretical Astrophysics (CITA), University of Toronto\\
                 email: {\tt fischera@cita.utoronto.ca} \\[\affilskip]
             $^4$Max Planck Institut f\"ur Astronomie, Heidelberg, Germany\\
                 email: {\tt brent@mpia.de} \\[\affilskip]
	     $^5$Jeremiah Horrocks Institute for Astrophysics and Supercomputing, University of        Central Lancashire, Preston, U.K.\\
                 email: {\tt cpopescu@uclan.ac.uk}}

\pubyear{2011} %% 
\volume{284}  %% insert here IAU Symposium No.
\pagerange{1--12}
\jname{The Spectral Energy Distribution of Galaxies}
\editors{R.J. Tuffs \&  C.C.Popescu, eds.}

\begin{document}

\maketitle

\begin{abstract}
We have carried out a detailed modeling of the dust Spectral Energy Distribution (SED) of the nearby, starbursting dwarf galaxy NGC 4214. A key point of our modeling is that we distinguish the emission from (i) HII regions and their associated photodissociation regions (PDRs) and (ii) diffuse dust. For both components we apply templates from the literature calculated with a realistic geometry and including radiation transfer. The large amount of existing data from the ultraviolet (UV) to the radio allows the direct measurement of most of the input parameters of the models. We achieve a good fit for the total dust SED of NGC 4214. In the present contribution we describe the available data, the data reduction and the determination of the model parameters, whereas a description of the general outline of our work is presented in the contribution of Lisenfeld et al. in this volume.

\end{abstract}

\firstsection % if your document starts with a section,
%    \vspace*{-0.2 cm}
\section{Introduction}

NGC 4214 is a nearby (2.94 Mpc; \cite{Apellaniz2002}) barred Magellanic irregular galaxy consisting of a smooth extended disk (Figure \ref{fig:SED}a) and a centrally concentrated young star forming (SF) region dominated by the two SF complexes NW and SE (Figure \ref{fig:SED}b). The metallicity of both complexes has been measured by \cite{Kobulnicky1996}, who found $Z\sim0.3Z_{\odot}$. The main stellar cluster in the NW complex, referred here as NW-A, is located at the center of the galaxy and has removed most of the gas in our line of sight. Using stellar synthesis models, \cite{Ubeda2007} determined the age, the mass, the radius and the extinction of the stellar clusters within the complexes NW and SE.

We have modeled the dust SED of NGC 4214 separately for the dust emission from SF regions and from the diffuse component. Due to its proximity, the large amount and good quality of the available data and previous studies of this object, most of the model parameters can be directly calculated, making this galaxy an excellent laboratory to improve our knowledge about dust in dwarf galaxies.

\section{Photometry}

\begin{figure}[ht]
% \vspace*{-2.0 cm}
\begin{center}
 \includegraphics[width=4.7in]{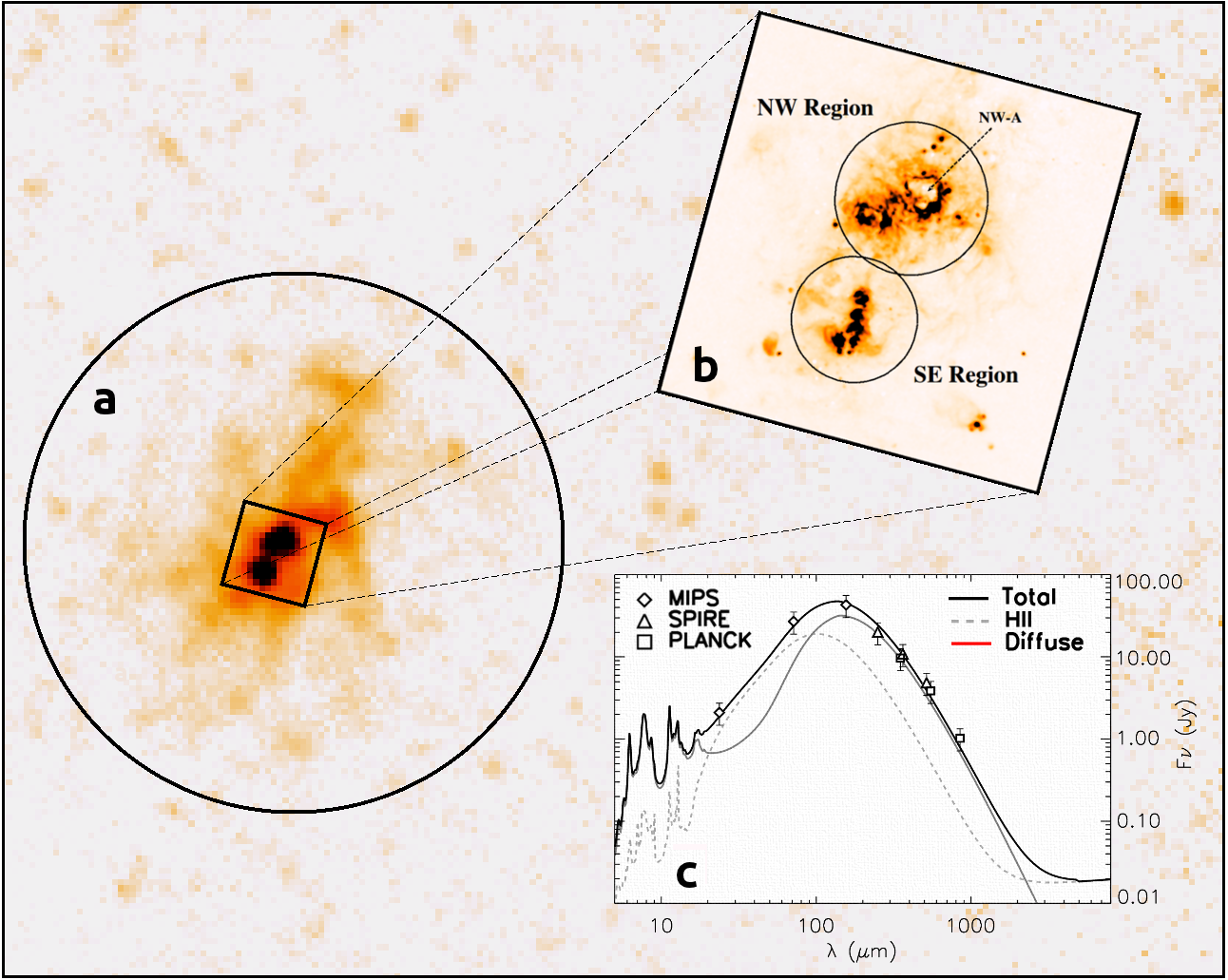} 
 \caption{\textbf{a)} SPIRE 250\,$\mu$m image of the disk of NGC 4214. The circle corresponds to the aperture we used to measure the global emission and it has a radius of $\sim$4\,kpc. \textbf{b)} HST-WFC3 H$\alpha$ map showing the complexes NW and SE and the location of the SSC NW-A. \textbf{c)} Best-fit model (black solid line) to the global SED for the parameters described in section \ref{sec:Parameters}. The global SED was obtained as the sum of the best-fits for the complexes NW and SE (grey dashed line) and the best-fit for the diffuse dust (grey solid line).}
   \label{fig:SED}
\end{center}

  \vspace*{-0.8 cm}

\end{figure}

In order to determine the observational dust SED of NGC 4214 we used data from \textit{Spitzer}, Herschel, and Planck archives, as well as our own observations at 1200\,$\mu$m using the MAMBO bolometer at the IRAM 30m telescope on Pico Veleta (Spain). For all the photometric measurements (i) color corrections and (ii) aperture corrections were applied and (iii) line decontamination was done. With the exception of the atomic line [CII] 158\,$\mu $m, which  contributes 4.26\% to the flux of MIPS 160\,$\mu$m, the contamination of the other considered lines ([OI] 63, [OIII] 88, [NII] 122,  [OI] 146 and [NII] 205\,$\mu $m as well as CO rotational transition lines) was found to be negligible. Figures \ref{fig:SED}a and \ref{fig:SED}b show the apertures used to measure the total emission and the emission from the SF complexes, respectively. In order to constrain the thermal radio emission we made use of the extinction corrected H$\alpha$ flux reported by \cite{MacKenty2000} and eq. (3) and (4a) in \cite{Condon1992}. The diffuse dust SED was determined by subtracting the combined emission of NW and SE complexes from the total emission of the galaxy.

   \vspace*{-0.6 cm}

\section{Determination of the parameters}
\label{sec:Parameters}

To model the \textbf{total dust emission} from the galaxy we used the model from \cite{Popescu2011} based on full radiation transfer calculations of the propagation of starlight in galaxy disks. The relevant parameters of this model in the case of NGC 4214 are (i) the radial scale-length of the old stellar disk, $h_s$, (ii) the central B-band face-on opacity, $\tau_{B}^{f}$, (iii) the star formation rate of the young stars, $SFR$, and (iv) the luminosity of the old stellar population, $OLD$. The diffuse SED, as well as the parameters $SFR$ and $OLD$, which control the diffuse radiation field, must be scaled by comparing the scale-length of NGC 4214 with the scale-length of the prototype galaxy NGC 891. In our case, these parameters can be observationally determined:  

\begin{itemize} 
 \item We adopted the value $\tau_{B}^{f}\sim1$ found by \cite{Ubeda2007} for the SSC NW-A.
 \item Following the instructions in \cite{Popescu2011} we determined $SFR$ and $OLD$ by integrating the observed UV-optical-NIR SED of NGC 4214. We obtained $SFR=0.15$\,M$_{\odot}/$yr whereas the luminosity of the old stellar population is negligible.
 \item We derived a scale-length in the B-band of $h_s\sim800$ pc by fitting elliptical isophotes.
 \end{itemize}

   \vspace*{0.1 cm}

To model the \textbf{dust emission from the HII regions} we used the model from \cite{Groves2008} based on radiation transfer calculations. This model computes the whole emergent spectrum including the attenuated stellar spectrum, the line emission spectrum, as well as a dust SED. The input parameters of this model are (i) the age of the stellar cluster, (ii) the metallicity, (iii) the hydrogen column density  of the surrounding PDR, $N_{\rm H}$, (iv) the external pressure, $p_{0}$, (v) the compactness, $C$, which parametrizes the dust grain temperature distribution, and (vi) the covering factor, $f_{\rm cov}$, which represents the fraction of the surface of the HII region which is covered by the PDR. With exception of $N_{\rm H}$, all the other parameters were observationally determined:

\begin{itemize}

  \item Following \cite{Ubeda2007}, we adopted 5 and 3.5 Myr for the age of the NW and SE
complex, respectively.

  \item We used a metallicity of Z$=0.2$Z$_{\odot}$ (results shown here) and also tested Z$=0.4$Z$_{\odot}$ which yielded very similar results (no template for Z$=0.3$Z$_{\odot}$ is available).
 
  \item We determined $p_{0}$ by comparing the expected and the observed radii of the HII regions as a function of the age. We found values of log$\left [(p _{0}/k) /\rm cm^{-3} K\right ]\sim6-7$.

  \item $C$ was directly determined from the external pressure and the mass of the clusters. We found values of 5.0 and 4.5 for the NW and SE complexes, respectively.

  \item In order to calculate $f _{cov}$ we assumed an optically thick,uniformly fragmented PDR surrounding the ionized gas. With these assumptions, the intrinsic luminosity of the central star cluster can be approximated by the sum of the observed luminosities of the stars, $L_{star}$, and the luminosity re-emitted by the dust, $L_{dust}$. In this picture, the covering factor is $f_{cov}=L_{dust}/(L_{dust}+L_{star})$. We obtained $f_{\rm cov}=0.44$ and 0.65 for NW and SE, respectively.

\end{itemize}

% \newpage

\section{Results}

Figure \ref{fig:SED}c shows the data points of the total observed dust SED, together with the model fits for the SF regions and for the diffuse emission, as well as the sum of both. A good fit could be achieved. Note that good fits were also obtained individually for the SF regions SE and NW (see Lisenfeld et al.). The fits were done with the input parameters determined from the observations, with the exception of $h_s$, for which we had to adopt a larger value (960-1260 pc) in order to obtain the correct dust temperature (see Lisenfeld et al. for more details).

\end{document}